# Nonlinear electromagnetic formulation for particle simulation of lower hybrid waves in toroidal geometry

J. Bao[1, 2], Z. Lin[2, *], A. Kuley[2], Z. X. Wang[2]

[1] Fusion Simulation Center, Peking University, Beijing 100871, China
[2] Department of Physics and Astronomy, University of California, Irvine, California 92697, USA

[*] Author to whom correspondence should be addressed. E-mail: *zhihongl@uci.edu*

Electromagnetic particle simulation model has been formulated and verified for nonlinear processes of lower hybrid (LH) waves in fusion plasmas. Electron dynamics is described by the drift kinetic equation using either kinetic momentum or canonical momentum. Ion dynamics is treated as the fluid system or by the Vlasov equation. Compressible magnetic perturbation is retained to simulate both the fast and slow LH waves. Numerical properties are greatly improved by using electron continuity equation to enforce consistency between electrostatic potential and vector potential, and by using the importance sampling technique. The simulation model has been implemented in the gyrokinetic toroidal code (GTC), and verified for the dispersion relation and nonlinear particle trapping of the electromagnetic LH waves.

## I. Introduction

Lower hybrid (LH) wave is one of the most efficient tool for steady state operation of tokamak to control current profile and suppress magnetohydrodynamics instabilities. Many experiments have demonstrated the success of lower hybrid current drive (LHCD)[1]. Thus, reliable prediction of current profile driven by LH wave is important for fusion experiments. Many linear and quasi-linear simulation models have been developed to study the LH wave propagation and absorption in tokamaks, such as Wentzel–Kramers–Brillouin (WKB) and full-wave approaches. [2-5] WKB method solves the Maxwell's equation in the short-wavelenth limit and gives the asymptotic solution for wave propagation, and full wave method solves the Maxwell's equation exactly in presence of a given particle distribution, however, these two approaches need to couple with a Fokker-Planck solver to address the absorption and driven current profile. Many important features of LH wave propagation and absorption in tokamak have been successfully explained based on the linear and quasi-linear models. For example, the "spectral gap"[6] phenomena (referring to the differences of LH parallel reflective index between launching location and absorption region) has been explained as the parallel spectrum up-shift due to toroidicity and broadening due to wave diffractions. However, the nonlinear effects of LH waves become more and more important in tokamak plasmas with the availability of high heating power.[1] For example, the unsolved problem: 'density limit'[7] (referring to the decrease of current drive efficiency at higher plasma density) is believed to be related to nonlinear parametric decay instability,[8] since the sideband waves have been observed in many experiments.[9-11] Particle-in-cell simulation approach is a powerful tool for studying nonlinear physics. Several algorithms and models of particle simulation of radio frequency (RF) waves in simple geometry (slab or cylinder) have been developed, e.g., GeFi[12-14], Vorpal[15,16] and G-gauge[17,18] codes. However, nonlinear RF simulation capability is unavailable for the toroidal geometry before our earlier work, in which we developed the electrostatic particle simulation of LH wave propagation in tokamaks.[19,20] In this work, we further extend the electrostatic particle model to a fully nonlinear electromagnetic particle model, which has been successfully implemented into the gyrokinetic toroidal code (GTC).[21] The electromagnetic dispersion relation and the nonlinear particle trapping of the LH waves have been verified by using current particle model.

In the model of this paper, ion dynamics is described by fluid equation for the LH wave propagation and absorption. However, ion kinetic effect is important in parametric processes related to LH waves, for example, nonlinear Landau damping will happen between some low frequency waves and ions, which has been shown in theory[22-25] and experiments.[9-11] In such cases, ion kinetic effect can be incorporated by using 6-D Vlasov equation.[19,26,27] Due to the fact that LH wave frequency is much smaller than electron cyclotron frequency $\omega \ll \Omega_{ce}$ and LH wavelength is much bigger than electron gyro-radius $k\rho_e \ll 1$, the electron dynamic is described by drift kinetic equation using either kinetic momentum or canonical momentum. At the same time, electron continuity equation is used for calculating the perturbed density from the perturbed current to avoid the numerical instabilities due to the inconsistency between electron density and current when calculated from the first two moments of the perturbed distribution function. To study the LH wave dynamics in tokamak core plasmas, the light wave and parallel electron plasma oscillation are removed in order to relax the constraints on the spatial grid size and time step size. The computational cost is huge for global simulation of LH waves in tokamak due to the short wavelength compared to the device size. For typical experimental parameters, LH wave frequency is on the order of GHz, the parallel reflective index is around 1~2, the perpendicular wavelength is the same order of magnitude as the electron skin depth, and the parallel wavelength is on the order of ion skin depth. Thus, resolving the short wavelength requires thousands of grids along radial, poloidal and toroidal directions. In order to minimize the computational requirements, we simulate the LH wave with a single toroidal mode number and utilize the toroidal symmetry by only simulating $2\pi/n$ length along the toroidal direction, where $n$ is the toroidal mode number of LH wave. This reduces the toroidal grid number from resolving hundreds of toroidal wavelength to one toroidal wavelength. However, in the future study of parametric process, the toroidal mode numbers of the decay waves are usually smaller than the pump LH wave, the toroidal length of the partial torus needs to be larger to resolve the lowest toroidal mode number. Furthermore, due to the fact that LH wave structure is usually localized in tokamaks, we do not need many markers in the region where LH wave amplitude is small. Thus, an importance sampling particle–in-cell (PIC) simulation scheme is implemented in order to utilize the markers with high efficiency.[28,29] We load more markers in the region where LH waves will propagate through, and use marker weight to represent the physical distribution. With above numerical techniques and simplifications on physics model, computational requirements are greatly reduced, while important features such as LH wave propagation, mode conversion and electron Landau damping are still captured to address the current drive problem.

In this paper, nonlinear formulation for LH waves is presented in Sec. II. Then comparison of the analytic dispersion relation between the particle model and the full Maxwell model in cold and uniform plasma limit is shown in Sec. III. The importance sampling particle-in-cell scheme in LH wave simulation is shown in Sec. IV. The verification of GTC simulation of LH wave dispersion relation and nonlinear particle trapping are shown in Sec. V. Sec. VI is the conclusion.

## II. Nonlinear formulation

For simulation of LH wave propagation and absorption with negligible damping from ion species, ion motion is described by fluid equation as described in Sec. II A. We provide two options for choosing the model of drift kinetic electron, namely, the electron model with canonical momentum and the electron model with kinetic momentum, and either of them can be used to study LH waves accurately with corresponding field equations which are given in Sec. II B and Sec. II C. Poisson's equation and electron perpendicular force balance equation are given in Sec. II D. In this paper, the notations $k_\parallel$ and $\mathbf{k}_\perp$ are the LH parallel and perpendicular wave vectors, respectively, with respect to the background magnetic field. $n_\parallel = ck_\parallel/\omega$ and $n_\perp = ck_\perp/\omega$ are the LH parallel and perpendicular reflective index, respectively. $\omega_{pe}$ is electron plasma frequency, $\omega_{pi}$ is the ion plasma frequency, $\Omega_{ce}$ is the electron cyclotron frequency, and $\Omega_{ci}$ is the ion cyclotron frequency.

## A. Fluid ion

Fluid ion dynamics is treated by the continuity and the momentum equations:

$$\frac{\partial \delta n_i}{\partial t} + \nabla \cdot \left[ \left( n_{i0} + \delta n_i \right) \boldsymbol{\delta u_i} \right] = 0, \tag{1}$$

$$m_i n_{i0} \frac{d\boldsymbol{\delta u_i}}{dt} = Z_i n_{i0} \left( \boldsymbol{\delta E} + \frac{1}{c} \boldsymbol{\delta u_i} \times \mathbf{B} \right) - \nabla \delta P_i, \tag{2}$$

where $\delta n_i$ and $\boldsymbol{\delta u_i}$ are the ion perturbed density and velocity, $n_{i0}$ is the ion equilibrium density, $Z_i$ and $m_i$ are the ion charge and mass, respectively. $\boldsymbol{\delta E} = -\nabla \phi - \frac{1}{c}\frac{\partial \boldsymbol{\delta A}}{\partial t}$ is the perturbed electric field. The total magnetic field is $\mathbf{B} = \mathbf{B_0} + \boldsymbol{\delta B}$, the equilibrium magnetic field is $\mathbf{B_0} = \nabla \times \mathbf{A_0}$ and the perturbed magnetic field is $\boldsymbol{\delta B} = \nabla \times \boldsymbol{\delta A}$. $\phi$ is the scalar potential, $\mathbf{A_0}$ is the equilibrium vector potential, and $\boldsymbol{\delta A}$ is the perturbed vector potential. $\delta P_i$ is the ion perturbed pressure, and it is approximated as $\delta P_i \approx \delta n_i T_{i0}$. In Eq. (2), $d/dt = \partial/\partial t + \boldsymbol{\delta u_i} \cdot \nabla$ is the total derivative including the convection term. For LH wave with $\omega \gg \mathbf{k}_\perp \cdot \mathbf{v_{thi}}$, we can drop the pressure term in Eq. (2) in the simulation, where $\mathbf{v_{thi}}$ is the ion thermal speed.

For the computational convenience of avoiding the calculation of $\partial \boldsymbol{\delta A}/\partial t$, we rewrite Eq. (2) to its canonical form:

$$m_i n_{i0} \frac{d\boldsymbol{\delta U_i}}{dt} = -Z_i n_{i0} \nabla \left[ \phi - \frac{1}{c} \boldsymbol{\delta u_i} \cdot (\mathbf{A_0} + \boldsymbol{\delta A}) \right], \tag{3}$$

where $\boldsymbol{\delta U_i} = \boldsymbol{\delta u_i} + \frac{Z_i}{cm_i}(\mathbf{A_0} + \boldsymbol{\delta A})$ is the canonical fluid velocity for ions. The total time derivative is defined as $d/dt = \partial/\partial t + \boldsymbol{\delta U_i} \cdot \nabla$. Eq. (3) is used in simulation instead of Eq. (2).

Furthermore, the 6-D Vlasov equation has been adopted for describing fully kinetic ion dynamics in GTC and can be utilized when ion kinetic effects are important.[19,26,27]

## B. Drift kinetic electron with canonical momentum

The canonical momentum formulation[30, 31] is proposed to remove the $\partial \delta A_\parallel / \partial t$ term in particle dynamic equations, which is difficult to implement as a time-centered finite difference in the simulation. In this section, we will introduce the implementation of the electron model with canonical momentum in our model.

The nonlinear drift kinetic equation with canonical momentum for electron is[30,31]:

$$L_c f_e (\mathbf{X}, p_\parallel, \mu, t) = 0, \tag{4}$$

where $f_e =$ is the electron distribution function and $L_c$ is the propagator in canonical form, $\mathbf{X}$, $p_\parallel$, $\mu$ and $t$ denote the particle position, canonical momentum $m_e v_\parallel + q_e \delta A_\parallel / c$, magnetic momentum and time, respectively.

The propagator $L_c$ consists of the equilibrium part $L_{c0}$, the first order perturbed part $\delta L_{c1}$ and the second order perturbed part $\delta L_{c2}$ as:

$$L_{c0} = \frac{\partial}{\partial t} + \left(\frac{p_\parallel}{m_e B_0}\mathbf{B_0^{*c}} + \frac{c\mu}{q_e B_0}\mathbf{b_0}\times\nabla B_0\right)\cdot\nabla - \frac{\mu}{B_0}\mathbf{B_0^{*c}}\cdot\nabla B_0 \frac{\partial}{\partial p_\parallel},$$

$$\delta L_{c1} = \left(-\frac{q_e\delta A_\parallel}{cm_e}\frac{\mathbf{B_0^{*c}}}{B_0} + \frac{c\mathbf{b_0}}{q_e B_0}\times q_e\nabla\Psi_l\right)\cdot\nabla - \frac{\mathbf{B_0^{*c}}}{B_0}\cdot q_e\nabla\Psi_l \frac{\partial}{\partial p_\parallel},$$

$$\delta L_{c2} = \frac{c\mathbf{b_0}}{q_e B_0}\times q_e\nabla\Psi_{nl}\cdot\nabla - \frac{\mathbf{B_0^{*c}}}{B_0}\cdot q_e\nabla\Psi_{nl}\frac{\partial}{\partial p_\parallel},$$

where $\mathbf{B_0} = B_0\mathbf{b_0}$ is the equilibrium magnetic field, $\mathbf{B_0^{*c}} = \mathbf{B_0} + \frac{c}{q_e}p_\parallel\nabla\times\mathbf{b_0}$. The generalized potential consists of the linear and nonlinear parts $\Psi = \Psi_l + \Psi_{nl}$, in which

$$\Psi_l = \phi - \frac{p_\parallel \delta A_\parallel}{m_e c} + \frac{\mu}{q_e}\delta B_\parallel \quad \text{and} \quad \Psi_{nl} = \frac{q_e \delta A_\parallel^2}{2m_e c^2}.$$

$\delta B_\parallel$ and $\delta A_\parallel$ are the compressional magnetic field perturbation and the parallel perturbed vector potential, respectively. The corresponding particle motion equations of Eq. (4) in magnetic coordinate are given in Appendix A.

The electron distribution function $f_e$ is also decomposed into the equilibrium and perturbed parts as $f_e = f_{e0} + \delta f_e$, and the equilibrium distribution function $f_{e0}$ satisfies the following equation:

$$L_{c0} f_{e0} = 0, \tag{5}$$

where we approximate $f_{e0}$ as a Maxwellian: $f_{e0} = n_{e0}\left(\frac{m_e}{2\pi T_{e0}}\right)^{3/2}\exp\left(-\frac{p_\parallel^2/m_e + 2\mu B}{2T_{e0}}\right)$, which neglects the neoclassical correction. In Eq. (5), the independent variable $p_\parallel$ reduces to $p_\parallel = m_e v_\parallel$ for the equilibrium part of the drift kinetic equation.

Taking into account of Eq. (5), we can rewrite Eq. (4) as:

$$L_{c0}\delta f_e + \delta L_{c1} f_{e0} + \delta L_{c1}\delta f_e + \delta L_{c2} f_{e0} + \delta L_{c2}\delta f_e = 0, \tag{6}$$

where the black part represents the first order linear term, the red and blue parts represent the second order and the third order nonlinear terms, respectively.

A perturbative $\delta f_e$ simulation method is applied to minimize the particle noise by defining the particle weight as $w_e = \delta f_e/f_e$, and the weight evolution equation can be written as:

$$\frac{dw_e}{dt} = L_c w_e = \frac{1}{f_e}L_c\delta f_e = -\frac{1}{f_e}(\delta L_{c1} + \delta L_{c2})f_{e0} = -(1-w_e)\frac{1}{f_{e0}}(\delta L_{c1} + \delta L_{c2})f_{e0}, \tag{7}$$

where the black and the red parts represent the linear and the nonlinear terms, respectively.

In principle, we can calculate both the density and parallel canonical velocity perturbations for calculating the perturbed fields from the kinetic particles. However, when we apply the $\delta f$ method and advance the weight equation by assuming $f_{e0} = f_{Maxwellian}$ as an exact Maxwellian in the calculation of $\nabla f_{e0}$ and $\partial f_{e0}/\partial p_\parallel$ in Eq. (7), the error from the marker noise between the marker distribution function $f_{Marker}$ and the exact Maxwellian will accumulate in Eq. (7), which is $\Delta f = f_{Marker} - f_{Maxwellian}$. Thus, after integrating the density and parallel canonical velocity perturbations from the marker distribution, the continuity equation will not be satisfied due to this error. The corresponding electrostatic potential and parallel vector potential will conflict with each other and cause numerical instabilities. This error can be reduced by increasing the marker number and will be eliminated when the marker number is infinite to build a perfect smooth Maxwellian as marker distribution function in the $\delta f$ simulation. In order to avoid this numerical issue, we use an additional electron continuity equation to time advance the electron density perturbation by the parallel canonical velocity perturbation calculated from the markers. The drift kinetic Eqs. (4) and (7) are only used for calculating the perturbed electron parallel canonical velocity and the perturbed pressure. This method provides a much better consistency between the scalar potential and vector potential in the $\delta f$ simulation since the continuity equation is satisfied all the time. There is a similar correction on the distribution function by neoclassical transport due to the discrepancy of the Maxwellian distribution and the physical neoclassical solution.[32] The comparison of a single LH mode excitation between with and without continuity equation is shown in Appendix B, which verifies the better numerical properties of using the continuity equation for the perturbed density.

Next, we integrate Eq. (6) for deriving the perturbed electron continuity equation, and keep the leading linear and nonlinear terms based on the orderings: $u_{\parallel e0} \approx 0$, $k_\parallel \ll k_\perp$, $ck_\perp/\omega_{pe} \sim 1$, $\delta n_e/n_{e0} \sim \delta P_\parallel/P_{\parallel 0} \sim \delta P_\perp/P_{\perp 0} \sim \delta B_\parallel/B_0 \sim |\delta \mathbf{B}_\perp|/B_0$, $\nabla n_{e0}/n_{e0} \sim 1/a$, $\nabla T_{e0}/T_{e0} \sim 1/a$, $\nabla B_0/B_0 \sim 1/R$, $a/R < 1$. The omitted terms are too small to affect the simulation results based on the orderings. Then we have:

$$\frac{\partial \delta n_e}{\partial t} + \underbrace{\mathbf{B_0} \cdot \nabla \left[ \frac{n_{e0}}{B_0} \left( \delta u_{\parallel ec} - \frac{q_e \delta A_\parallel}{m_e c} \right) \right]}_{\{I\}} + \underbrace{B_0 \mathbf{v_E} \cdot \nabla \left( \frac{n_{e0}}{B_0} \right)}_{\{II\}} - \underbrace{n_{e0} \left( \mathbf{v}^* + \mathbf{v_E} \right) \cdot \frac{\nabla B_0}{B_0}}_{\{III\}}$$

$$-\underbrace{\frac{c\mathbf{b_0} \times \nabla P_{\perp 0}}{q_e B_0} \cdot \frac{\nabla \delta B_\parallel}{B_0}}_{\{IV\}} + \underbrace{B_0 \mathbf{v_E} \cdot \nabla \left( \frac{\delta n_e}{B_0} \right)}_{\{V\}} - \underbrace{\frac{c\mathbf{b_0} \times \nabla \delta P_\perp}{q_e B_0} \cdot \frac{\nabla \delta B_\parallel}{B_0}}_{\{VI\}} - \underbrace{\mathbf{B_0} \cdot \nabla \left( \frac{\delta n_e}{B_0} \frac{q_e \delta A_\parallel}{m_e c} \right)}_{\{VII\}}, \quad (8)$$

$$+ \underbrace{\nabla \times \left( \delta A_\parallel \mathbf{b_0} \right) \cdot \nabla \left[ \frac{n_{e0}}{B_0} \left( -\frac{c}{4\pi n_{e0} q_e} \nabla_\perp^2 \delta A_\parallel + \delta u_{\parallel i} \right) \right]}_{\{VIII\}} = 0$$

where $\mathbf{v_E} = \dfrac{c\mathbf{b_0} \times \nabla \phi}{B_0}$ is the $\mathbf{E} \times \mathbf{B}$ drift, $\mathbf{v}^* = \dfrac{\mathbf{b_0} \times \nabla(\delta P_\perp + \delta P_\parallel)}{n_{e0} m_e \Omega_{ce}}$ is the diamagnetic drift with $\delta P_\parallel = \dfrac{1}{m_e} \int \mathbf{dv} \, p_\parallel^2 \delta f_e$ and $\delta P_\perp = \int \mathbf{dv} \, \mu B_0 \delta f_e$, $\delta u_{\parallel ec} = \dfrac{1}{n_{e0} m_e} \int \mathbf{dv} \, p_\parallel \delta f_e$ is the canonical velocity, and $\int \mathbf{dv} = \dfrac{2\pi B_0}{m_e^2} \int dp_\parallel d\mu$. The perpendicular equilibrium pressure in term {IV} is defined as $P_{\perp 0} = \int \mathbf{dv} \, \mu B_0 f_{e0} = n_{e0} T_{e0}$. The black part represents the linear terms, the red part represents the nonlinear terms. The term {I} is the linear parallel compressional term, terms {II}-{IV} represent the linear work by the leading drifts, term {V} represents the $\mathbf{E} \times \mathbf{B}$ nonlinearity, term {VI} represents the diamagnetic drift nonlinearity, {VII} is the parallel nonlinear term and {VIII} is the nonlinear magnetic compressional term.

Here, we use parallel Ampere's law for solving $\delta A_\parallel$ as:

$$\left( \nabla_\perp^2 - \dfrac{\omega_{pe}^2}{c^2} \right) \delta A_\parallel = -\dfrac{4\pi}{c} \left( J_{\parallel i} + J_{\parallel e} \right), \tag{9}$$

where $J_{\parallel i} = Z_i n_{i0} \delta u_{\parallel i}$ and $J_{\parallel e} = \dfrac{q_e}{m_e} \int \delta f_e p_\parallel \mathbf{dv}$, $\delta u_{\parallel i}$ is the parallel mechanical fluid velocity of ion. The second term on the LHS of Eq. (9) arises due to the difference between $p_\parallel$ and $m_e v_\parallel$.

Inverting the Ampere's law Eq. (9), we have the relation: $\delta u_{\parallel ec} - \dfrac{q_e \delta A_\parallel}{m_e c} = -\dfrac{c}{4\pi n_{e0} q_e} \nabla_\perp^2 \delta A_\parallel + \delta u_{\parallel i}$, which is used in term {I} of Eq. (8) for a better numerical stability.

## C. Drift kinetic electron with kinetic momentum

Although the canonical momentum formulation has some computational advantages, the kinetic momentum formulation (also called symplectic formulation) is more transparent regarding the physical meaning of each term in the equations[31]. In this section, we introduce the kinetic momentum formulation as an alternative electron model.

Using guiding center position $\mathbf{X}$, parallel velocity $v_\parallel$ and magnetic momentum $\mu$ as independent variables in five dimensional phase space, drift kinetic Vlasov equation for electron is[31,33]:

$$L_k f_e(\mathbf{X}, v_\parallel, \mu, t) = 0, \tag{10}$$

where $f_e(\mathbf{X}, v_\parallel, \mu, t)$ is the electron distribution function and $L_k$ is the propagator in sympletic form. $L_k$ can be decomposed into the equilibrium part $L_{k0}$, the first order perturbed part $\delta L_{k1}$ and the second order perturbed part $\delta L_{k2}$ as $L_k = L_{k0} + \delta L_k$, with

$$L_{k0} = \frac{\partial}{\partial t} + \left( \frac{v_\parallel}{B_0} \mathbf{B_0^{*k}} + \frac{c\mathbf{b_0}}{q_e B_0} \times \mu \nabla B_0 \right) \cdot \nabla - \frac{\mu}{m_e B_0} \mathbf{B_0^{*k}} \cdot \nabla B_0 \frac{\partial}{\partial v_\parallel},$$

$$\delta L_{k1} = \left[ v_\parallel \frac{\delta \mathbf{B}_\perp}{B_0} + \frac{c\mathbf{b_0}}{q_e B_0} \times \left( q_e \nabla \phi + \mu \nabla \delta B_\parallel \right) \right] \cdot \nabla$$
$$+ \left[ -\frac{\mu}{m_e B_0} \delta \mathbf{B}_\perp \cdot \nabla B_0 - \frac{\mathbf{B_0^{*k}}}{m_e B_0} \cdot \left( q_e \nabla \phi + \mu \nabla \delta B_\parallel \right) - \frac{q_e}{cm_e} \frac{\partial \delta A_\parallel}{\partial t} \right] \frac{\partial}{\partial v_\parallel},$$

$$\delta L_{k2} = -\frac{\delta \mathbf{B}_\perp}{m_e B_0} \cdot \left( q_e \nabla \phi + \mu \nabla \delta B_\parallel \right) \frac{\partial}{\partial v_\parallel},$$

where $\mathbf{B^{*k}} = \mathbf{B_0^{*k}} + \delta \mathbf{B}_\perp$, $\mathbf{B_0^{*k}} = \mathbf{B_0} + \frac{B_0 v_\parallel}{\Omega_{ce}} \nabla \times \mathbf{b_0}$ and $\delta \mathbf{B}_\perp = \nabla_\perp \times \left( \delta A_\parallel \mathbf{b_0} \right)$. The corresponding particle motion equations of Eq. (10) in magnetic coordinate are given in Appendix A.

Next, we will do the same procedure in Sec. II B to deduce the weight evolution equation and electron continuity equation for electron model with kinetic momentum. The distribution function is decomposed into the equilibrium and perturbed part as $f_e = f_{e0} + \delta f_e$. The equilibrium distribution $f_{e0}$ obeys the following equation:

$$L_{k0} f_{e0} = 0, \tag{11}$$

where $f_{e0}$ is also approximated as a Maxwellian: $f_{e0} = n_{e0} \left( \frac{m_e}{2\pi T_{e0}} \right)^{3/2} \exp\left( -\frac{m_e v_\parallel^2 + 2\mu B}{2 T_{e0}} \right)$.

From Eqs. (10) and (11), we have:

$$L_{k0} \delta f_e + \delta L_{k1} f_{e0} + \delta L_{k1} \delta f_e + \delta L_{k2} f_{e0} + \delta L_{k2} \delta f_e = 0, \tag{12}$$

where the black part represents the first order linear term, the red and blue parts represent the second order and the third order nonlinear terms, respectively.

Defining the particle weight as $w_e = \delta f_e / f_e$, and the weight evolution equation is:

$$\frac{dw_e}{dt} = L_k w_e = \frac{1}{f_e} L_k \delta f_e = -\frac{1}{f_e} \left( \delta L_{k1} + \delta L_{k2} \right) f_{e0} = -(1 - w_e) \frac{1}{f_{e0}} \left( \delta L_{k1} + \delta L_{k2} \right) f_{e0}, \tag{13}$$

where the red part represents the nonlinear term.

In the electron model with kinetic momentum, we also use electron continuity equation for numerical stability as discussed in Sec. II B. Integrating Eq. (13) in velocity space and keeping the leading linear and nonlinear terms based on the same orderings in Sec. II B, we get the electron continuity equation as:

$$\frac{\partial \delta n_e}{\partial t} + \mathbf{B_0} \cdot \nabla \left( \frac{n_{e0} \delta u_{\|e}}{B_0} \right) + B_0 \mathbf{v_E} \cdot \nabla \left( \frac{n_{e0}}{B_0} \right) - n_{e0} \left( \mathbf{v}^* + \mathbf{v_E} \right) \cdot \frac{\nabla B_0}{B_0} + \frac{c \mathbf{b_0} \times \nabla \delta B_\|}{B_0^2} \cdot \frac{\nabla P_{\perp 0}}{q_e}$$
$$+ \textcolor{red}{\delta \mathbf{B_\perp} \cdot \nabla \left( \frac{n_{e0} \delta u_{\|e}}{B_0} \right) + B_0 \mathbf{v_E} \cdot \nabla \left( \frac{\delta n_e}{B_0} \right) + \frac{c \mathbf{b_0} \times \nabla \delta B_\|}{B_0^2} \cdot \frac{\nabla \delta P_\perp}{q_e}} = 0 \quad (14)$$

where $\delta u_{\|e} = \frac{1}{n_{e0}} \int v_\| \delta f_e \, d\mathbf{v}$ is the perturbed parallel velocity, $\mathbf{v_E} = \frac{c \mathbf{b_0} \times \nabla \phi}{B_0}$ is the $\mathbf{E} \times \mathbf{B}$ drift velocity, $\mathbf{v}^* = \frac{\mathbf{b_0} \times \nabla (\delta P_\perp + \delta P_\|)}{n_{e0} m_e \Omega_{ce}}$ is the perturbed diamagnetic drift velocity with $\delta P_\| = m_e \int d\mathbf{v} \, v_\|^2 \delta f_e$ and $\delta P_\perp = \int d\mathbf{v} \, \mu B_0 \delta f_e$, the perpendicular equilibrium pressure is defined as $P_{\perp 0} = \int d\mathbf{v} \, \mu B_0 f_{e0} = n_{e0} T_{e0}$, and $\int d\mathbf{v} = \frac{2\pi B_0}{m_e} \int dv_\| d\mu$. The terms related to $\delta B_\|$ in Eq. (14) are the diamagnetic drift due to the parallel perturbed magnetic field, the other terms are the same with Eq. (28) in Ref. 34. The black part represents the linear terms and the red part represents the nonlinear terms in Eq. (14).

In order to derive $\partial \delta A_\| / \partial t$ term for pushing the particles, firstly we take the time derivative of the parallel Ampere's law:

$$\nabla_\perp^2 \frac{\partial \delta A_\|}{\partial t} = -\frac{4\pi}{c} \left( Z_i n_{i0} \frac{\partial \delta u_{\|i}}{\partial t} + q_e n_{e0} \frac{\partial \delta u_{\|e}}{\partial t} \right). \quad (15)$$

Secondly, we integrate Eq. (12) to get the momentum equation based on the same ordering with Eq. (14) as:

$$n_{e0} \frac{\partial \delta u_{\|e}}{\partial t} + \frac{q_e n_{e0}}{m_e} \mathbf{b_0} \cdot \nabla \phi + \frac{q_e n_{e0}}{m_e c} \frac{\partial \delta A_\|}{\partial t} + \frac{B_0}{m_e} \mathbf{b_0} \cdot \nabla \left( \frac{\delta P_\|}{B_0} \right) + \frac{\mathbf{B_0} \cdot \nabla \delta B_\|}{m_e B_0^2} P_{\perp 0}$$
$$+ \textcolor{red}{\frac{q_e \delta n_e}{m_e} \mathbf{b_0} \cdot \nabla \phi + \frac{q_e \delta n_e}{m_e c} \frac{\partial \delta A_\|}{\partial t} + \frac{q_e n_{e0}}{m_e B_0} \delta \mathbf{B_\perp} \cdot \nabla \phi + \frac{\delta \mathbf{B_\perp}}{m_e} \cdot \nabla \left( \frac{\delta P_\|}{B_0} \right) + B_0 \mathbf{v_E} \cdot \nabla \left( \frac{n_{e0} \delta u_{\|e}}{B_0} \right)} = 0 \quad (16)$$

where the black part represents the linear terms and the red part represents the second order nonlinear terms.

Thirdly, $\partial \delta u_{\|i} / \partial t$ can be calculated from Eq. (2) as:

$$\frac{\partial \delta u_{\|i}}{\partial t} = \frac{Z_i}{m_i} \left( -\mathbf{b_0} \cdot \nabla \phi - \frac{1}{c} \frac{\partial \delta A_\|}{\partial t} \right). \quad (17)$$

Substituting Eqs. (16) and (17) into Eq. (15), we can derive the following equation for $\partial \delta A_\parallel / \partial t$:

$$\left(\nabla_\perp^2 - \frac{\omega_{pe}^2}{c^2} - \frac{\omega_{pi}^2}{c^2}\right)\frac{\partial \delta A_\parallel}{\partial t} = \frac{\omega_{pe}^2}{c^2}\xi_\parallel, \tag{18}$$

where

$$\xi_\parallel = c\left(1+\frac{m_e}{m_i}\right)\mathbf{b_0}\cdot\nabla\phi + \frac{c}{n_{e0}q_e}\mathbf{B_0}\cdot\nabla\left(\frac{\delta P_\parallel}{B_0}\right) + \frac{c}{n_{e0}q_e}\frac{\mathbf{B_0}\cdot\nabla\delta B_\parallel}{B_0^2}P_{\perp 0} + c\frac{\delta n_e}{n_{e0}}\mathbf{b_0}\cdot\nabla\phi$$
$$+ \frac{c}{B_0}\delta\mathbf{B_\perp}\cdot\nabla\phi + \frac{c}{n_{e0}q_e}\delta\mathbf{B_\perp}\cdot\nabla\left(\frac{\delta P_\parallel}{B_0}\right) + \frac{cm_e B_0}{n_{e0}q_e}\mathbf{v_E}\cdot\nabla\left(\frac{n_{e0}\delta u_{\parallel e}}{B_0}\right) + \frac{\delta n_e}{n_{e0}}\frac{\partial \delta A_\parallel}{\partial t}.$$

Eq. (18) with nonlinear terms is solved by an iterative method, we solve it without last term in $\xi_\parallel$ firstly, then substitute the result of $\partial \delta A_\parallel / \partial t$ into $\xi_\parallel$ and solve Eq. (18) again to get the new $\partial \delta A_\parallel / \partial t$. One iteration is enough to resolve the small quantity $(\delta n_e/n_{e0})(\partial \delta A_\parallel/\partial t)$ in $\xi_\parallel$ since it convergent very fast. On the other hand, we can also move $(\delta n_e/n_{e0})(\partial \delta A_\parallel/\partial t)$ in $\xi_\parallel$ to the LHS of Eq. (18), and solve $\partial \delta A_\parallel / \partial t$ directly without iteration. However, this requires to build the matrix $\nabla_\perp^2 - \frac{\omega_{pe}^2}{c^2} - \frac{\omega_{pi}^2}{c^2} - \frac{\delta\omega_{pe}^2}{c^2}$ (where $\delta\omega_{pe}^2 = \frac{4\pi\delta n_e e^2}{m_e}$) at each time step in the simulation, which will slow down the computational speed.

In Sec. II B and Sec. II C, electron models with canonical momentum and with kinetic momentum have been discussed, respectively. Both of them apply the continuity equation to avoid the numerical instability. The leading terms of the continuity equations from this two models are the same except for an additional term {VII} in Eq. (8), which is due to the difference between $p_\parallel$ and $v_\parallel$ as an independent variable of these two models. However, the different term is very small compared to the sum of other terms, and either of these two models can be applied in the simulation with the corresponding field equations.

## D. Field equations

The Poisson's equation for fluid (or fully kinetic) ion and drift kinetic electron is[31,35]:

$$\nabla_\perp^2\phi + \underbrace{\nabla_\perp\cdot\left(\frac{\omega_{pe}^2}{\Omega_{ce}^2}\nabla_\perp\phi\right)}_{I} = -4\pi\left(Z_i\delta n_i + q_e\delta n_e\right) + 4\pi\nabla_\perp\cdot\left(\underbrace{\frac{q_e n_{e0}}{B_0^2}\mathbf{B_0}\times\delta\mathbf{A_\perp}}_{II} + \underbrace{\frac{cm_e n_{e0}u_{\parallel e0}}{B_0^2}\nabla_\perp\delta A_\parallel}_{III}\right) \tag{19}$$

In Eq. (19), the parallel electron plasma oscillation is suppressed by assuming $\nabla^2 \approx \nabla_\perp^2$ in the first term on the LHS. Term {I} is the electron polarization density caused by the polarization drift, term {II} is from the inductive part $(\mathbf{b_0}/B)\times\partial\boldsymbol{\delta A}_\perp/\partial t$ of electron $\mathbf{E}\times\mathbf{B}$ motion, and term {III} is the electron polarization density caused by the magnetic-flutter motion along perturbed magnetic-field lines. $u_{\|e0} = (1/n_{e0})\int v_\| f_{e0}\mathbf{dv}$ in term {III} is the electron equilibrium flow.

In LH wave frequency range $\omega \ll \Omega_{ce}$, we can use electron perpendicular force balance equation to solve $\delta B_\|$ and $\phi$ together with Eq. (19), which is given as following:

$$n_e q_e \boldsymbol{\delta E}_\perp = \nabla_\perp \cdot \boldsymbol{\delta P_e} - \frac{1}{c}\mathbf{J}_{e\perp} \times \mathbf{B_0}, \tag{20}$$

where the perpendicular electric field $\boldsymbol{\delta E}_\perp$ is defined as:

$$\boldsymbol{\delta E}_\perp = -\nabla_\perp \phi - \frac{1}{c}\frac{\partial \boldsymbol{\delta A}_\perp}{\partial t}, \tag{21}$$

and the divergence of the electron pressure is

$$\nabla_\perp \cdot \boldsymbol{\delta P_e} \approx \nabla_\perp \left(\delta P_\perp + \frac{n_{e0} T_{e0} \delta B_\|}{B_0}\right), \tag{22}$$

and the electron perpendicular current is

$$\mathbf{J}_{e\perp} \approx \frac{c}{4\pi}\left[\nabla_\perp \delta B_\| \times \mathbf{b_0} + \nabla_\| \times (\nabla_\perp \delta A_\| \times \mathbf{b_0})\right] - \mathbf{J}_{i\perp}. \tag{23}$$

In Eq. (22), we neglect the electron polarization drift $\mathbf{v_{pol}} = (q_e/m_e \Omega_{ce}^2)(\partial \boldsymbol{\delta E}_\perp/\partial t)$ contribution to the pressure term based on the drift kinetic electron assumption (the electron Larmor radius is much smaller than the perpendicular wavelength $k_\perp \rho_e \approx 0$), since the pressure caused by the polarization drift is $\delta P_{pol} \approx \delta n_{pol} T_{e0} = q_e n_{e0} \rho_e^2 \nabla_\perp^2 \phi \approx 0$. The first term on the RHS is calculated from the guiding center dynamics, and the second term on the RHS is from the inductive part of electron $\mathbf{E}\times\mathbf{B}$ motion: $(\mathbf{b_0}/B)\times\partial\boldsymbol{\delta A}_\perp/\partial t$, which does not appear in the guiding center dynamic equation explicitly.

Taking the perpendicular divergence operation on both sides of Eq. (20), we have:

$$-\nabla_\perp^2 \phi + \frac{1}{c}\mathbf{b_0}\cdot\nabla\frac{\partial \delta A_\|}{\partial t} = \nabla_\perp \cdot\left[\frac{1}{n_{e0}q_e}\nabla_\perp\left(\delta P_\perp + \frac{n_{e0} T_{e0} \delta B_\|}{B_0}\right)\right] \\ + \nabla_\perp \cdot\left(\frac{B_0}{4\pi n_{e0}q_e}\nabla_\perp \delta B_\|\right) + \frac{1}{n_{e0}q_e c}\nabla_\perp\cdot(\mathbf{J}_{i\perp}\times\mathbf{B_0}) \tag{24}$$

Although in LH frequency range we have $|\mathbf{J}_{i\perp}|/|\mathbf{J}_{e\perp}| \approx (m_e/m_i)(\Omega_{ce}/\omega_{LH}) \ll 1$, where $\omega_{LH} \approx \sqrt{\Omega_{ce}\Omega_{ci}}$, we keep $\mathbf{J}_{i\perp}$ related terms in Eq. (24) for the correctness of the simulation in the lower frequency range $\omega \ll \omega_{LH}$. Here, the Coulomb gauge is described as: $\nabla_\perp \cdot \boldsymbol{\delta A}_\perp + \nabla_\parallel \cdot (\delta A_\parallel \mathbf{b_0}) = 0$, so the second term on the LHS of Eq. (24) comes from the relation $\nabla_\perp \cdot (\partial \boldsymbol{\delta A}_\perp/\partial t) = -\mathbf{b_0} \cdot \nabla (\partial \delta A_\parallel/\partial t)$. This term is much smaller than the first term on the LHS in LH wave simulation and thus can be dropped.

In the multi-pass cases, when the LH wave propagates to the edge region, the reflections of LH wave will happen at the cutoffs where $\omega = \omega_{pe}$, and the perpendicular reflective index $n_\perp$ will decrease to zero very quickly. Thus, the equilibrium density scale length is comparable to the wave length $k_\perp L_n \sim 1$ near the cutoffs in the edge region, the terms related to the non-uniformity of the equilibrium need to be kept in equations (19) and (24). However, the cutoff region with $\omega = \omega_{pe}$ is removed in this model by using the approximation $\nabla^2 \approx \nabla_\perp^2$ to the first term on the LHS of Poisson's equation Eq. (19). Thus, current model can not address the reflection of LH waves at the cutoffs. In this paper, we focus on the single-pass study of LH wave in the core plasmas, namely, most energy of the LH wave can be absorbed before reaching the cutoffs near the plasma edge. In the core plasmas, the wavelength of LH wave is much smaller than the equilibrium plasma scale length $L_0 \sim \left(L_n = \frac{2\pi n_{e0}}{\nabla n_{e0}}, L_T = \frac{2\pi T_{e0}}{\nabla T_{e0}}, L_B = \frac{2\pi B_0}{\nabla B_0}\right)$, namely, $k_\perp L_0 \gg 1$ can be guaranteed during the simulation. Furthermore, we can assume the electron equilibrium flow $u_{\parallel e0} = 0$ for a Maxwellian distribution of electron. Thus, we can simplify Eq. (19) and Eq. (24) as:

$$\left(1 + \frac{\omega_{pe}^2}{\Omega_{ce}^2}\right)\nabla_\perp^2 \phi + \frac{4\pi n_{e0} q_e}{B_0}\delta B_\parallel = -4\pi(Z_i \delta n_i + q_e \delta n_e), \tag{25}$$

$$\delta B_\parallel = \frac{4\pi}{B_0(1 + 0.5\beta_e)}(n_{e0} q_e \chi - n_{e0} q_e \phi - \delta P_\perp), \tag{26}$$

where $\beta_e = 8\pi n_{e0} T_{e0}/B_0^2$, and $\chi$ can be derived from the following equation:

$$\nabla_\perp^2 \chi = -\frac{1}{n_{e0} q_e c}\nabla_\perp \cdot (\mathbf{J}_{i\perp} \times \mathbf{B_0}). \tag{27}$$

From Eq. (26), we notice that in fluid ion (or fully kinetic ion[12]) and drift kinetic (DK) electron model, the perturbed force balance for $\delta B_\parallel$ is different from the gyrokinetic (GK) ion and DK electron model for low frequency modes, which is $\delta B_\parallel + 4\pi\delta P_\perp/B_0 = 0$.[31] This is due to the fact that electron $\mathbf{E}\times\mathbf{B}$ motion can not cancel with ion species in LH frequency range. Substituting Eq. (26) into Eq. (25), we can solve the following equation to derive $\phi$:

$$\left(1+\frac{\omega_{pe}^2}{\Omega_{ce}^2}\right)\nabla_\perp^2\phi - \frac{\omega_{pe}^2}{\Omega_{ce}^2}\frac{\omega_{pe}^2}{c^2}\frac{\phi}{1+0.5\beta_e} = -4\pi\left[Z_i\delta n_i + q_e\delta n_e - q_e\frac{\beta_e}{(2+\beta_e)}\frac{\delta P_\perp}{T_{e0}}\right] \\ -\frac{\omega_{pe}^2}{\Omega_{ce}^2}\frac{\omega_{pe}^2}{c^2}\frac{\chi}{1+0.5\beta_e}. \tag{28}$$

$\mathbf{\delta A_\perp}$ can be solved from:

$$\nabla_\perp^2\mathbf{\delta A_\perp} = -\nabla_\perp\delta B_\parallel \times \mathbf{b_0}. \tag{29}$$

Now, Eqs. (1), (3-4), (7-9) and (26-29) form a closed system for electron model with canonical momentum, while Eqs. (1), (3), (10), (13-14), (18) and (26-29) form a closed system for electron model with kinetic momentum. In linear and nonlinear regimes, both electron models with kinetic and canonical momentum can be used in LH wave studies with similar complexity and numerical performance.

## III. The analytic dispersion relation from electromagnetic particle model

In order to verify the validity of the electromagnetic models given in section II, we derive the corresponding linear dispersion relation and compare with the result from Maxwell equations in the limit of uniform cold plasmas.

We start from the electron model with kinetic momentum, and its corresponding Eqs. (1), (3), (10), (13-14), (18) and (26-29) are used.

In cold and uniform plasmas, Eq. (18) reduces to:

$$\left(\nabla_\perp^2 - \frac{\omega_{pe}^2}{c^2} - \frac{\omega_{pi}^2}{c^2}\right)\frac{\partial\delta A_\parallel}{\partial t} = \frac{\omega_{pe}^2}{c}\left(1+\frac{m_e}{m_i}\right)\mathbf{b_0}\cdot\nabla\phi, \tag{30}$$

Eq. (26) reduces to:

$$\delta B_\parallel = -\frac{\omega_{pe}^2}{\Omega_{ce}c}(\phi-\chi), \tag{31}$$

and Eq. (28) reduces to:

$$\left(1+\frac{\omega_{pe}^2}{\Omega_{ce}^2}\right)\nabla_\perp^2\phi - \frac{\omega_{pe}^2}{\Omega_{ce}^2}\frac{\omega_{pe}^2}{c^2}\phi = -4\pi(Z_i\delta n_i + q_e\delta n_e) - \frac{\omega_{pe}^2}{\Omega_{ce}^2}\frac{\omega_{pe}^2}{c^2}\chi. \tag{32}$$

The ion dynamics is described by Eq. (3) in canonical form, which has the numerical advantage by avoiding calculating $\partial \delta \mathbf{A}_\perp / \partial t$. For the convenience of theoretical analysis, we use the equivalent Eq. (2) in cold plasma limit and decompose it into parallel and perpendicular linear components:

$$\delta u_{i\|} = \frac{iZ_i}{m_i \omega} \delta E_\| \tag{33}$$

$$\boldsymbol{\delta u_{i\perp}} = \frac{iZ_i \omega}{m_i \left( \omega^2 - \Omega_{ci}^2 \right)} \left[ \boldsymbol{\delta E_\perp} + \frac{i\Omega_{ci}}{\omega} \boldsymbol{\delta E_\perp} \times \mathbf{b_0} \right]. \tag{34}$$

After linearization, the ion continuity Eq. (1) in uniform plasmas can be written as:

$$\frac{\partial \delta n_i}{\partial t} + n_{i0} \nabla \cdot \boldsymbol{\delta u_i} = 0 \tag{35}$$

Electron dynamic is described by the continuity equation and drift kinetic equation. In uniform and cold plasmas, electron continuity Eq. (14) reduces to:

$$\frac{\partial \delta n_e}{\partial t} + n_{e0} \mathbf{b_0} \cdot \nabla \delta u_{e\|} = 0. \tag{36}$$

Integrating the drift kinetic Eq. (12) (equivalent to Eqs. (10) and (13)) for the momentum moment in cold and uniform plasma limit:

$$\frac{\partial \delta u_{e\|}}{\partial t} + \frac{q_e}{m_e} \mathbf{b_0} \cdot \nabla \phi + \frac{q_e}{m_e c} \frac{\partial \delta A_\|}{\partial t} = 0. \tag{37}$$

From Eqs. (30)-(37) with applying the Fourier transform: $\partial_t \to -i\omega$, $\mathbf{b_0} \cdot \nabla \to ik_\|$, and $\nabla_\perp \to i\mathbf{k}_\perp$, we can get the linear dispersion relation in cold and uniform plasmas as:

$$S + D^2 \frac{1}{n_\perp^2 - S'} = -(P-1) \frac{n_\|^2}{n_\perp^2 - (P-1)}, \tag{38}$$

where $S$, $P$ and $D$ are the elements of the cold plasma dielectric tensor in Stix notation with frequency $\omega \ll \Omega_{ce}$ as following:

$$S = 1 + \frac{\omega_{pe}^2}{\Omega_{ce}^2} - \frac{\omega_{pi}^2}{\omega^2 - \Omega_{ci}^2},$$

$$P = 1 - \frac{\omega_{pe}^2}{\omega^2} - \frac{\omega_{pi}^2}{\omega^2},$$

$$D = -\frac{\omega_{pe}^2}{\omega \Omega_{ce}} + \frac{\omega_{pi}^2 \Omega_{ci}}{\omega \left( \omega^2 - \Omega_{ci}^2 \right)}.$$

And $S'$ in Eq. (38) is given as:

$$S' = S - 1 - \frac{\omega_{pe}^2}{\Omega_{ce}^2} = -\frac{\omega_{pi}^2}{\omega^2 - \Omega_{ci}^2}.$$

Eq. (38) is the dispersion relation derived by using electron model with kinetic momentum. We can also get the same result by using the electron model with canonical momentum, which consists of Eqs. (1), (3-4), (7-9) and (26-29).

We rewrite Eq. (38) into the determinant form and compare with the well-known result from Maxwell model[36] in Table I:

**TABLE I.** The analytic dispersion relations derived from the reduced model and the Maxwell model, respectively.

| Reduced model solution | Maxwell model solution |
|---|---|
| $\begin{vmatrix} S - n_\parallel^2 & -iD & n_\parallel n_\perp \\ iD & S' - n_\perp^2 & 0 \\ n_\parallel n_\perp & 0 & P - 1 - n_\perp^2 \end{vmatrix} = 0$ | $\begin{vmatrix} S - n_\parallel^2 & -iD & n_\parallel n_\perp \\ iD & S - n_\perp^2 - n_\parallel^2 & 0 \\ n_\parallel n_\perp & 0 & P - n_\perp^2 \end{vmatrix} = 0$ |

Compared to the Maxwell model solution, the difference in $S'$ of our reduced model is due to the fact that we drop the displacement current and polarization current in the perpendicular electron force balance equation. The vacuum term is lost in the parallel diagonal term $P - 1 - n_\perp^2$ of the reduced model, since we remove the electron plasma wave by dropping the $\nabla_\parallel^2 \phi$ term in Poisson's equation, and remove the light wave by dropping the displacement current in Ampere's law. The reason why $-n_\parallel^2$ term in the second diagonal term is missing in the reduced model is due to the fact that we drop some coupling terms of parallel and perpendicular components in Ampere's law and force balance equation by assuming $|\nabla_\perp| \gg |\nabla_\parallel|$.

Thus, our simulation model is accurate for the waves in the core region of typical tokamak where the plasma density is high such that $\omega \ll \omega_{pe}$ or $|P| \gg 1$ and the wave's perpendicular reflective index is much larger than the parallel reflective index $(n_\perp^2 \gg n_\parallel^2)$. Namely, our simulation results can recover the Maxwell model results when $|S' - n_\perp^2| \gg |1 + \omega_{pe}^2/\Omega_{ce}^2 - n_\parallel^2|$ and $|P| \gg 1$ are satisfied simultaneously. The high frequency light wave and electron plasma wave are artificially removed, so it enables us to use the bigger space grid size and time step size without resolving the high frequency waves with short wavelength. This is sufficient and efficient to the single-pass of LH waves without cutoff. However, our simulation fails when LH waves propagate to the cutoff layer in the plasma edge where $n_\perp^2 \sim 0$ and the electron plasma wave and light wave effects can not be ignored. Thus, current field model for core plasma need to couple with the fully Maxwell model for edge plasma in order to address the multi-pass physics accurately.

# IV. Importance sampling for particle-in-cell simulation

In order to reduce the numerical noise and the computational cost in marker particle simulation, it is helpful to load many markers at the initial time in the region where the LH wave propagates through, while a very small number of markers are required in the other region where LH wave perturbation is small. Thus, the importance sampling techniques is applied to particle-in-cell simulation[29] of the LH waves. Here, we give an example of this scheme based on the electron model with canonical momentum. The marker distribution is defined as $g_e(\mathbf{X}, p_\parallel, \mu, t) = g_{e0}(\mathbf{X}, p_\parallel, \mu) + \delta g_e(\mathbf{X}, p_\parallel, \mu, t)$, where $g_{e0} = g_e(t=0)$ is the initial sampling marker distribution, and $\delta g_e$ is the perturbed marker distribution.

Similar to Eq. (4), the drift kinetic equation for maker distribution can be written as:

$$L_c g_e = (L_{c0} + \delta L_{c1} + \delta L_{c2})(g_{e0} + \delta g_e) = 0. \tag{39}$$

Instead of a single weight as defined in Sec. II, two weights are used in generalized weight-based scheme. The total weight is defined as

$$p_e = \frac{f_e}{g_e}, \tag{40}$$

which represents the importance of each marker to $f_e$. And the perturbed weight is defined as

$$w_e = \frac{\delta f_e}{g_e}, \tag{41}$$

which represents the importance of each marker to $\delta f_e$.

Thus, considering Eqs. (4), (6) and (39), the total weight evolution equation is:

$$\frac{dp_e}{dt} = 0, \tag{42}$$

and the perturbed weight evolution equation becomes:

$$\frac{dw_e}{dt} = -(p_e - w_e)\frac{1}{f_{e0}}(\delta L_{c1} + \delta L_{c2})f_{e0}. \tag{43}$$

Eqs. (42) and (43) determine the evolution of the total distribution $f_e$ and the perturbed distribution $\delta f_e$, respectively. Because the marker distribution does not need to be proportional to the physical distribution in the importance sampling scheme: $g_e \neq C \bullet f_e$, where $C$ is a constant, we need to evolve Eqs. (42) and (43) with considering the importance of the markers. Furthermore, we can also apply this scheme on electron model with kinetic momentum by using $\delta L_{k1} + \delta L_{k2}$ instead of $\delta L_{c1} + \delta L_{c2}$ in Eq. (43).

In principle, we can sample arbitrary $g_{e0}(\mathbf{X}, p_\|, \mu)$ initially in order to achieve the local high resolution in phase space where $\delta f$ amplitude is high. For the volume conservation in phase space as shown by Eq. (39), the perturbed marker distribution will evolve through the following equation:

$$L_c \delta g_e = -(\delta L_{c1} + \delta L_{c2}) g_{e0} - L_{c0} g_{e0}. \tag{44}$$

The reason to keep the last term on the RHS of Eq. (44) is that $L_{c0} g_{e0} \neq 0$ in general when we choose an approximate $g_{e0}$ for optimal phase space sampling. Finite $\delta g_e$ makes $g_e$ different from the initial arrangement $g_{e0}$, and changes the desired numerical resolution. However, the time scale for the marker evolution is much longer than LH wave period $|L_c g_e / g_e| \ll \omega$. Thus, the desired numerical resolution does not vary much for the duration of LH wave simulation.

The general magnetic flux coordinate system $(\psi, \theta, \zeta)$ is used for the simulations of LH waves in toroidal geometry, where $\psi$ is the poloidal flux function, $\theta$ is the magnetic poloidal angle and $\zeta$ is the magnetic toroidal angle. For the LH wave propagation case with launching from $\theta = 0$, we sample many more markers in the region where the LH wave will propagate as shown in Fig. 1. The coordinates (X, Z) in Figs. 1, 2(a) and 2(d) represent the horizontal and vertical distances measured from the geometric center of the tokamak and the color scale in Fig. 1 represents the number of the makers per cell used in the simulation. In the simulation, the axis values of the electron plasma temperature and the plasma density are $T_{e0} = 1.0 keV$ and $n_{i0} = n_{e0} = 5.0 \times 10^{13} cm^{-3}$, respectively. We choose the other parameters based on the orders of magnitude of the Alcator C tokamak, which includes $a = 0.16m$, $R_0 = 0.64m$ and the axis value of the magnetic field $B_a = 5.0T$. The launched LH wave frequency $f_0 = 4.6GHz$ and the toroidal refractive index $n_t = ck_t / \omega = 1.86$. The comparison of the numerical performance between uniform sampling and non-uniform sampling is shown in Fig. 2. We find that the mode structures in the poloidal plane and the flux-surfaces with non-uniform sampling are much smoother than the uniform sampling. The horizontal coordinate $\alpha = \theta - \zeta/q$ in Figs. 2(c) and 2(f) is the magnetic field line label. The horizontal coordinate $m$ is defined as poloidal mode number in Figs. 2(b) and 2(e). The high $m$ poloidal components of the wave-packet are nearly zero in the importance sampling case as shown in Fig. 2(b), while the high $m$ poloidal components of the wave-packet have larger amplitudes in ordinary sampling case as shown in Fig. 2(e), which proves that the importance sampling particle-in-cell method helps to decrease the numerical noise and suppress the numerical high $k_\theta$ modes.

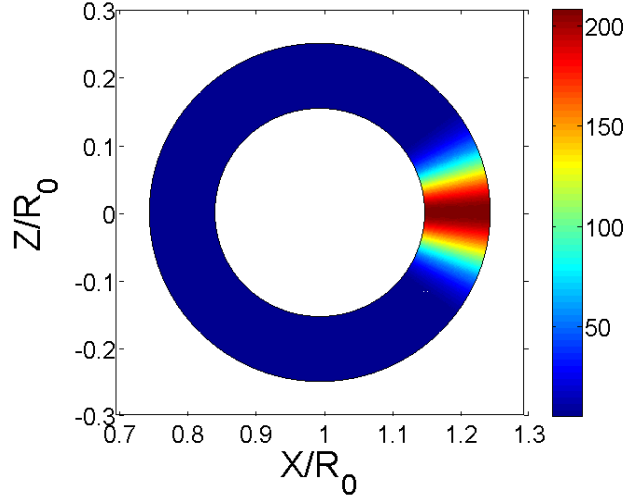

**FIG. 1.** The marker distribution in real space of the importance sampling in the simulation of LH wave propagation. The color scale represents the number of the makers per cell.

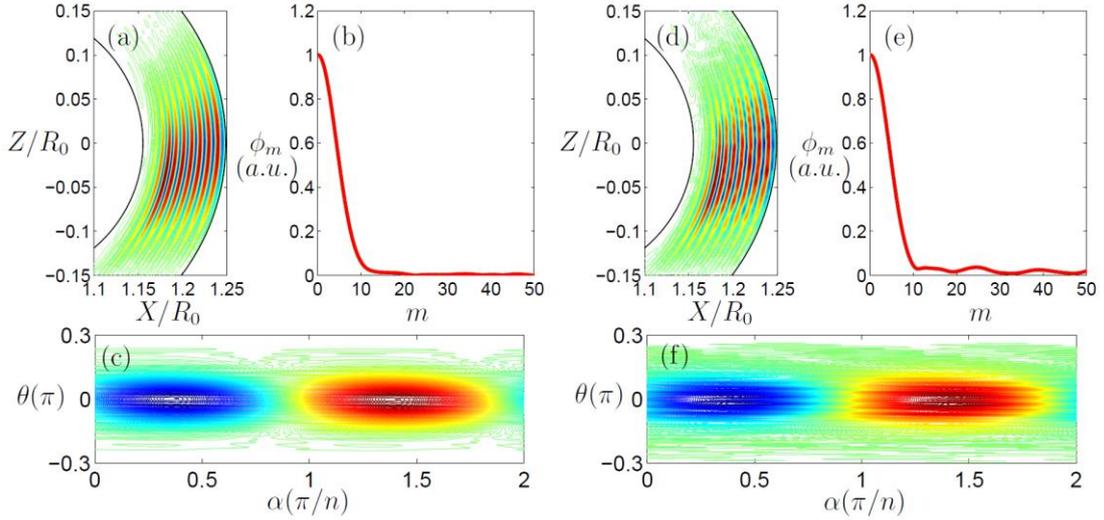

**FIG. 2.** Panels (a), (b) and (c) are from the importance sampling PIC simulation, which show the LH wave structure in poloidal plane, the poloidal spectrum of the wave-packet and the LH wave structure on flux-surface, respectively. Panels (d), (e) and (f) are from the ordinary PIC simulation. The color scale in panels (a), (c), (d) and (f) represents the electrostatic potential $\phi$ in a. u. unit. $m$ represents the poloidal harmonic number in panels (b) and (e). The mode structures in the poloidal plane and the flux surfaces with importance sampling as shown by (a) and (c) are much smoother than the conventional sampling as shown by (d) and (f). The amplitude of the numerical high $m$ harmonics in the importance sampling (b) are much smaller than the conventional sampling (e).

# V. Verification of GTC simulation of dispersion relation and nonlinear particle trapping of LH waves

In this section, we will show the dispersion relation benchmark between our simulation model and analytical theory. The model described in Section 2 has been implemented in the gyrokinetic toroidal code (GTC). GTC[37] has been successfully applied to simulate microturbulence,[38] energetic particle transport,[39] Alfven eigenmodes,[40,41] and magnetohydrodynamic instabilities including kink mode[42] and tearing mode[43] in fusion plasmas. In order to derive the theoretical solution for benchmarks, the simulations are performed in the cylinder geometry of GTC with uniform magnetic field in this section. GTC simulations of LH waves for different $k_\perp/k_\parallel$ regimes are carried out by using initial perturbation method. For these benchmark cases, plasma density $n_{e0} = n_{i0} = 2 \times 10^{13} cm^{-3}$, electron temperature $T_{e0} = 50.0 eV$ (for cold plasma) and magnetic field $B = 2.0T$ are uniform, and the magnetic field is only along the axial direction in cylinder. The parallel wave vector in cylinder is $k_\parallel = n/R = 100.0 m^{-1}$, where $n = 100$ is the parallel mode number and $R = 1.0m$ (the length of the cylinder is $l = 2\pi R$), and the radius is $a = 0.3m$. The simulations are carried by using both canonical momentum and kinetic momentum electron model, and their comparison with theoretical results are shown in Fig. 3. It is seen that there are two branches of waves from Fig. 3: the slow wave and the fast wave. The perpendicular phase velocity $v_{p\perp} = \omega/k_\perp$ and group velocity $v_{g\perp} = \partial\omega/\partial k_\perp$ have the same sign for the fast wave, which corresponds to the LHS part of the dispersion relation curve in Fig. 3, while they have opposite signs for the slow wave, which corresponds to the RHS part of the dispersion relation curve in Fig. 3. Simulation results agree with the reduced model and the Maxwell model solutions very well when $k_\perp/k_\parallel \gg 1$. With typical experimental parameters, $k_\perp/k_\parallel \gg 1$ can be satisfied for LH waves in the core plasmas.[46,47]

Next, we use initial perturbation method[44,45] to carry out the electromagnetic simulations of the linear Landau damping and nonlinear Landau damping for LH waves in hot plasmas. In the initial perturbation method, an electron density perturbation with $k_\parallel = 150.0 m^{-1}$ is initiated, and then the perturbation evolves self-consistently. The plasma density $n_{e0} = n_{i0} = 7.6 \times 10^{13} cm^{-3}$, electron temperature $T_{e0} = 6.0 keV$ and magnetic field $B = 2.0T$ are uniform. In this parameter regime, both electrostatic and electromagnetic components are important to LH wave dispersion relation. The time histories of the generalized potential $\psi$ (as defined in Sec. II) of LH waves with $\omega/(k_\parallel v_{the}) \approx 3.2$ from the linear and nonlinear electromagnetic simulations are shown in Fig.4 (a), in which the red solid line shows an oscillation in amplitude in the nonlinear simulation, while the blue dashed line shows the wave decays exponentially in the linear simulation. The oscillation of the LH wave amplitude in nonlinear simulation is due to the wave trapping of resonant electrons, and the oscillation (bounce) frequency agrees well with theoretical prediction $\omega_b = k_\parallel v_{the} \sqrt{e\psi/T_{e0}}$ as shown in Fig 4(b). The particle trapping by waves is a basic phenomenon of the nonlinear wave-particle interaction[50, 51]. The agreement between simulation and theory for the bounce frequency shows that our model captures the important nonlinear effects faithfully.

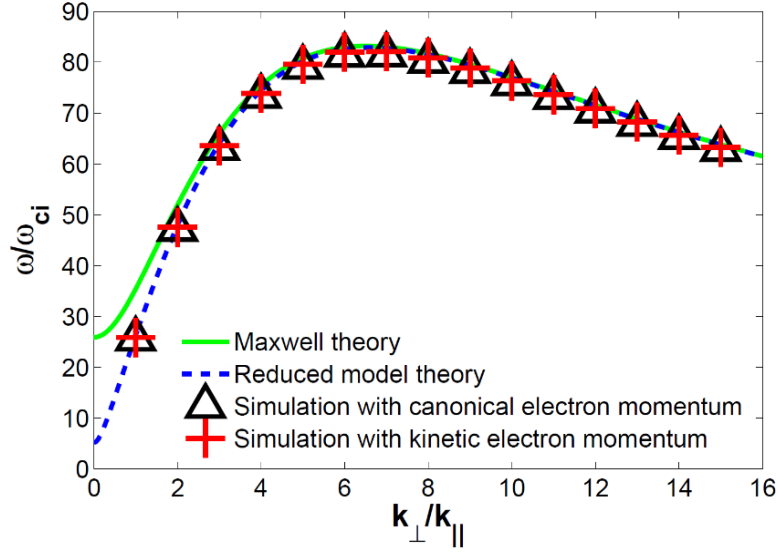

**FIG. 3.** Comparison of the electromagnetic dispersion relation of LH wave between GTC simulation and analytic theory.

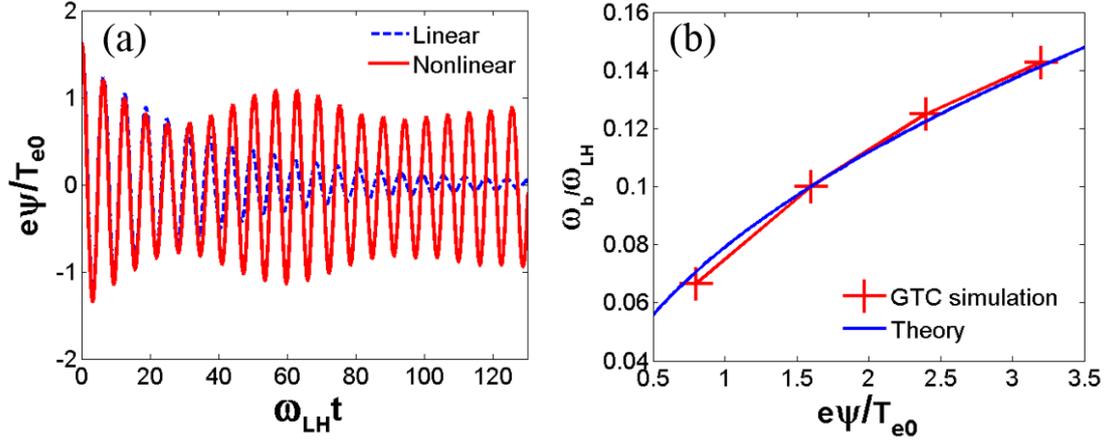

**FIG. 4.** (a)Nonlinear simulation of LH wave exhibits oscillation in amplitude of electrostatic potential (solid line), while linear simulation shows exponential decay (dashed line). (b)The comparison of bounce frequency dependence on wave amplitude between GTC nonlinear simulation and theory.

# VI. Conclusions

The nonlinear and electromagnetic fluid (or fully kinetic) ion/drift kinetic electron model has been implemented into GTC for LH wave study in the toroidal geometry. Drift kinetic electron is described by both canonical momentum and kinetic momentum, and either of the descriptions can be applied in simulation. The use of electron continuity equation provides a better numerical performance, which avoids the discrepancy between marker distribution with noise and the Maxwellian distribution in the $\delta f$ simulation. Both the analytic results from the model and numerical results from GTC have been verified with the dispersion relation of LH waves. In the nonlinear simulation of LH wave damping in hot plasmas, we find that the amplitude of the wave field perturbation oscillates with a bounce frequency, which is due to the wave trapping of resonant electrons. An importance sampling particle-in-cell scheme has been applied to simulate LH wave propagation with the high numerical resolution and efficiency. Compared to WKB and full-wave approaches based on the linear and quasi-linear theory, our PIC simulation model based on the first-principles can capture the nonlinear effects, such as particle trapping by waves, which provides a powerful tool to study the nonlinear physics of LH waves in tokamak. Applications of this simulation model to the linear mode conversion, nonlinear current drive and parametric decay instabilities of LH waves will be reported in separate papers.[48, 52]

# Acknowledgments

This work was supported by China National Magnetic Confinement Fusion Science Program (Grant No. 2013GB111000) and US Department of Energy (DOE) SciDAC GSEP Program. This work used resources of the Oak Ridge Leadership Computing Facility at Oak Ridge National Laboratory (DOE Contract No. DE-AC05-00OR22725) and the National Energy Research Scientific Computing Center (DOE Contract No. DE-AC02-05CH11231). JB acknowledges support from China Scholarship Council (Grant No. 201306010032), and thanks encouragements from and useful discussions with L. Chen, N. J. Fisch, M. Porkolab, W. W. Lee, F. Zonca, J. C. Wright, Z. X. Lu, N. Xiang, D. H. Li and GTC team.

# Appendix A: Drift kinetic electron motion equation in magnetic coordinate with fully electromagnetic perturbations

The general magnetic flux coordinate system $(\psi, \theta, \zeta)$ has already been defined in Sec. IV. Then the equilibrium magnetic field can be written either in contravariant form as Eq. (A1) or in covariant form as Eq. (A2):

$$\mathbf{B}_0 = q\nabla\psi \times \nabla\theta - \nabla\psi \times \nabla\zeta, \tag{A1}$$

$$\mathbf{B}_0 = \delta\nabla\psi + I\nabla\theta + g\nabla\zeta. \tag{A2}$$

The Jacobian in magnetic flux coordinates is

$$J^{-1} = \nabla\psi \cdot \nabla\theta \times \nabla\zeta = \frac{B_0^2}{gq+I}, \tag{A3}$$

Writing the particle motion equations in Eqs. (4) and (10) into magnetic flux coordinates, we have:

$$\dot{\psi} = \frac{1}{q_e}\frac{\partial \varepsilon_0}{\partial B_0}\left(\frac{I}{D}\frac{\partial B_0}{\partial \zeta} - \frac{g}{D}\frac{\partial B_0}{\partial \theta}\right) + \frac{I}{D}\frac{\partial \phi}{\partial \zeta} - \frac{g}{D}\frac{\partial \phi}{\partial \theta}$$
$$+ v_\parallel B_0\left(\frac{g}{D}\frac{\partial \alpha}{\partial \theta} - \frac{I}{D}\frac{\partial \alpha}{\partial \zeta}\right) + \frac{\mu}{q_e}\left(\frac{I}{D}\frac{\partial \delta B_\parallel}{\partial \zeta} - \frac{g}{D}\frac{\partial \delta B_\parallel}{\partial \theta}\right), \tag{A4}$$

$$\dot{\theta} = \frac{v_\parallel B_0(1-\rho_c g' - g\partial_\psi \alpha)}{D} + \frac{g}{D}\left(\frac{1}{q_e}\frac{\partial \varepsilon_0}{\partial B_0}\frac{\partial B_0}{\partial \psi} + \frac{\partial \phi}{\partial \psi} + \frac{\mu}{q_e}\frac{\partial \delta B_\parallel}{\partial \psi}\right), \tag{A5}$$

$$\dot{\zeta} = \frac{v_\parallel B_0(q+\rho_c I' + I\partial_\psi \alpha)}{D} - \frac{I}{D}\left(\frac{1}{q_e}\frac{\partial \varepsilon_0}{\partial B_0}\frac{\partial B_0}{\partial \psi} + \frac{\partial \phi}{\partial \psi} + \frac{\mu}{q_e}\frac{\partial \delta B_\parallel}{\partial \psi}\right), \tag{A6}$$

$$\dot{\rho}_\parallel = -\frac{(1-\rho_c g' - g\partial_\psi \alpha)}{D}\left(\frac{1}{q_e}\frac{\partial \varepsilon}{\partial B_0}\frac{\partial B_0}{\partial \theta} + \frac{\partial \phi}{\partial \theta} + \frac{\mu}{q_e}\frac{\partial \delta B_\parallel}{\partial \theta}\right)$$
$$-\frac{(q+\rho_c I' + I\partial_\psi \alpha)}{D}\left(\frac{1}{q_e}\frac{\partial \varepsilon}{\partial B_0}\frac{\partial B_0}{\partial \zeta} + \frac{\partial \phi}{\partial \zeta} + \frac{\mu}{q_e}\frac{\partial \delta B_\parallel}{\partial \zeta}\right), \tag{A7}$$
$$+\frac{(I\partial_\zeta \alpha - g\partial_\theta \alpha)}{D}\left(\frac{1}{q_e}\frac{\partial \varepsilon}{\partial B_0}\frac{\partial B_0}{\partial \psi} + \frac{\partial \phi}{\partial \psi} + \frac{\mu}{q_e}\frac{\partial \delta B_\parallel}{\partial \psi}\right) - \frac{\partial \alpha}{\partial t}$$

$$\dot{\rho}_c = -\frac{(1-\rho_c g')}{D}\left[\frac{1}{q_e}\frac{\partial \varepsilon}{\partial B_0}\frac{\partial B_0}{\partial \theta} + \frac{\partial \phi}{\partial \theta} + \frac{\mu}{q_e}\frac{\partial \delta B_\parallel}{\partial \theta} - \frac{q_e}{m_e}\rho_\parallel B_0^2\frac{\partial \alpha}{\partial \theta}\right]$$
$$-\frac{(q+\rho_c I')}{D}\left(\frac{\partial \phi}{\partial \zeta} + \frac{\mu}{q_e}\frac{\partial \delta B_\parallel}{\partial \zeta} - \frac{q_e}{m_e}\rho_\parallel B_0^2\frac{\partial \alpha}{\partial \zeta}\right), \tag{A8}$$

where $D = JB_0^2(1+\rho_c \mathbf{b_0}\cdot \nabla \times \mathbf{b_0}) = gq + I + \rho_c(gI' - Ig')$, $I' = \partial I/\partial \psi$, $g' = \partial g/\partial \psi$,

$\alpha = \delta A_\parallel/B_0$, $\rho_\parallel = v_\parallel/\Omega_{ce}$, $\rho_c = \rho_\parallel + \alpha$ and $\frac{\partial \varepsilon_0}{\partial B_0} = \mu + \frac{q_e^2}{m_e}\rho_\parallel^2 B_0$. Eqs. (A4)-(A8) describe

the drift kinetic electron dynamics with fully electromagnetic perturbations in magnetic flux coordinate, and they can reduce to the results from R. B. White et al.[49] when $\delta B_\parallel = 0$. Eqs. (A4)-(A6) and (A8) describe the particle motion for the drift kinetic electron with canonical momentum Eq. (4), and Eqs. (A4)-(A7) describe the particle motion for the drift kinetic electron with kinetic momentum Eq. (10).

# Appendix B: Comparison of numerical properties with and without the electron continuity equation

Here, we carry out antenna excitation of the single LH mode simulation in order to test the numerical performance of the cases with and without electron continuity equation. In the simulation, the plasma equilibrium parameters are the same with the dispersion relation benchmark case in Sec. V. The LH mode with frequency $\omega = 80.0\Omega_{ci}$ and parallel wave vector $k_{\parallel} = 100.0 m^{-1}$ is chosen. In the first simulation, we use electron continuity equation to calculate the electron perturbed density and show the mode and amplitude histories in Figs. 5(a) and 5(b). It is found that the mode history has a good linear growth. In the second simulation, we use the kinetic marker to calculate the electron perturbed density in the simulation. We find that the real and imaginary parts do not match with each other, and the mode amplitude history has a large numerical oscillation with the same marker number as shown in Figs. 5(c) and 5(d). Only after increasing the marker number in the third simulation, the real and imaginary parts match with each other better as the first case, which is shown in Figs. 5(e) and 5(f).

By comparing these three cases, we find that applying electron continuity equation can help to suppress the numerical instability and reduce the computational cost as illustrated in Sec. II.

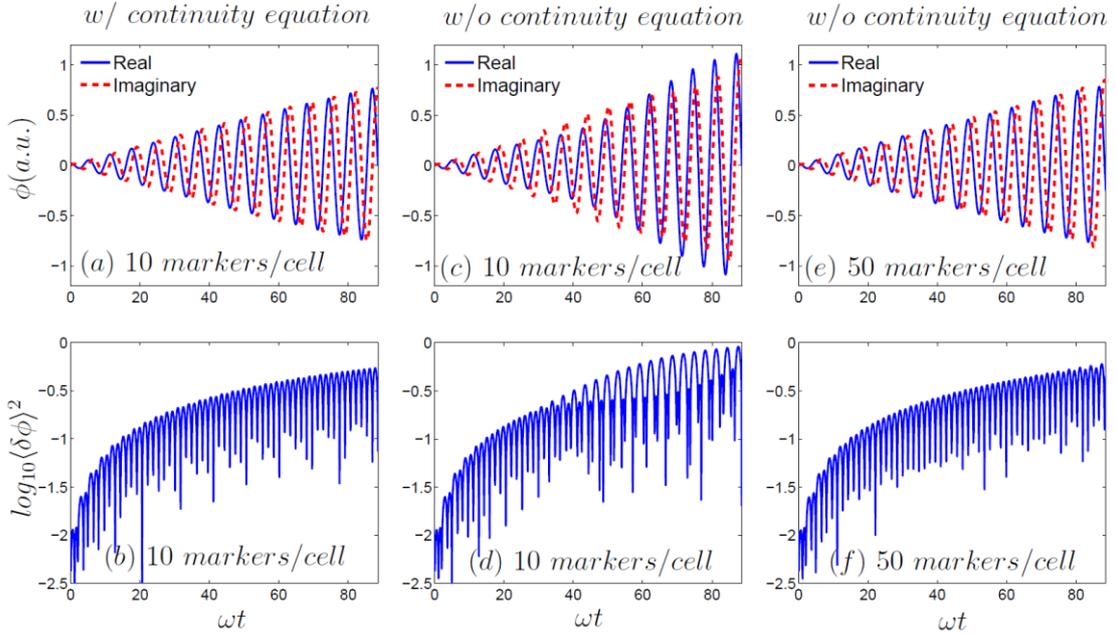

**FIG. 5.** Panels (a), (c) and (e) show the single mode history of electrostatic potential, and panels (b), (d) and (f) show the amplitude history. Panels (a) and (b) are from the case with 10 markers per cell and with using electron continuity equation. Panel (c) and (d) are from the case with 10 markers per cell and without using electron continuity equation. Panel (e) and (f) are from the case with 50 markers per cell and without using electron continuity equation.